\documentclass{aa}
\usepackage{txfonts}
\usepackage{natbib}
\usepackage{graphicx}
\newcommand{\Al}{\element[][26\!]{Al}\ }
\newcommand{\Aln}{\element[][26\!]{Al}}
\begin{document}

\title{Line shape diagnostics of Galactic $^{26\!}$Al}

\author{K. Kretschmer\inst{1}  
  \and R. Diehl\inst{1} 
  \and D.~H. Hartmann\inst{2}
}
\institute{
  Max-Planck-Institut f\"ur extraterrestrische Physik,
  Postfach 1312, D-85741 Garching, Germany 
  \and
  Department of Physics \& Astronomy, Clemson University, Clemson, SC
  29634-0978, USA
}

\offprints{K. Kretschmer,
\email{kkr@mpe.mpg.de}}
\date{Received 4 July 2003 / Accepted 6 November 2003}

% \authorrunning{K. Kretschmer, R. Diehl \& D. H. Hartmann}
% \titlerunning{\Al in the Galaxy}

\abstract { The shape of the gamma-ray line from radioactive \Aln, at
  1808.7~keV energy in the frame of the decaying isotope, is
  determined by its kinematics when it decays, typically 10$^6$~y after
  its ejection into the interstellar medium from its nucleosynthesis
  source.  Three measurements of the line width exist: HEAO-C's 1982
  value of $(0+3)$~keV FWHM, the GRIS 1996 value of $(5.4\pm 1.3)$~keV
  FWHM, and the recent RHESSI value of $(2.0\pm 0.8)$~keV FWHM,
  suggesting either ``cold'', ``hot'', or ``warm'' \Al in the ISM.  We model
  the line width as expected from Galactic rotation, expanding
  supernova ejecta, and/or Wolf-Rayet winds, and predict a value below
  1~keV (FWHM) with plausible assumptions about \Al initial velocities
  and expansion history.  Even though the recent RHESSI measurement
  reduces the need to explain a broad line corresponding to
  540~km~s$^{-1}$ mean \Al velocity through extreme assumptions about
  grain transport of \Al or huge interstellar cavities, our results
  suggest that standard \Al ejection models produce a line on the narrow 
  side of what is observed by RHESSI and INTEGRAL. Improved INTEGRAL and RHESSI
  spatially-resolved line width measurements should help to disentangle the
  effects of Galactic rotation from the ISM trajectories of \Aln.
  \keywords{nuclear reactions, nucleosynthesis, abundances -- gamma
    rays: observations -- supernovae: general -- ISM: supernova
    remnants -- stars: formation.}}

\maketitle

\section{Introduction}

Current Galactic nucleosynthesis reveals itself through the decay of
\Aln, one of its radioactive by-products with a mean lifetime of
$1.04\times 10^6$~y.  \Al undergoes $\beta^+$-decay into an excited
state of $^{26}$Mg, which
de-excites through emission of a gamma-ray photon at 1808.7 keV. This
gamma-ray line has been observed and imaged throughout the Galaxy
\citep{mahoney82,diehl95,oberlack97,knoedlseder99,plueschke01}.
Sources of \Al may be AGB stars and novae, but massive stars (via
core-collapse supernovae and winds from Wolf-Rayet stars) have been
found the most 
plausible and probably dominating sources \citep{prantzos96}. The
rather irregular \Al emission along the plane of the Galaxy, and its
consistency with the patterns of tracers of massive-star activity,
is the main argument for favouring massive stars as the
sources \citep{diehl96,knoedlseder99,plueschke01}.  Flux measurements
have been employed to study the nature of the sources, comparing with
predicted \Al yields from models of the source types. The amount of \Al
present in the ISM of the Galaxy has been estimated at $\approx
2~\mathrm{M}_\odot$, and used to argue for the roles of different
source types. But the
uncertainties about the spatial distribution and total number of
nucleosynthesis events add to source yield uncertainties, 
providing only qualitative arguments for the nature of \Al sources
\citep{prantzos96}. Therefore, locally constrained candidate source
populations have been studied, such as in the Cygnus region
\citep{knoedlseder00, plueschke01}. In such a case, the distance to
the sources is constrained to a smaller interval, along with the
radial velocity due to Galactic rotation.

Measurements of Galactic \Al with high-resolution spectrometers have
produced somewhat controversial results: The initial discovery with
the space-borne \mbox{HEAO-C} Ge spectrometer had reported a narrow line
\citep[intrinsic width (FWHM) $(0+3)$~keV;][]{mahoney84}. But the GRIS
balloon-borne Ge spectrometer found the line to be significantly
broadened (intrinsic FWHM $(5.4\pm1.4)$~keV; \citet{naya96}).  Such a
broad line, however, cannot be explained easily \citep[see][]{chen97}.
If the origin of the line broadening was thermal, the \Al decay region
would need to be at a temperature of $\approx 4.5\times 10^8$~K.
Alternatively, \Al isotopes would have to maintain a mean velocity
around 540 km~s$^{-1}$ over their 1 My decay time, travelling kpc
distances at those speeds. One may either assume that a substantial
fraction of \Al is injected into such rather large interstellar
cavities, or that a substantial fraction of \Al condenses onto dust
grains before deceleration, so it maintains its momentum throughout
passages of supernova remnant shells or other obstacles.  The velocity
could also be a result of re-acceleration of dust grains by
interstellar shocks in the neighbourhood of the source, allowing them
to maintain a high velocity over the \Al decay time scale
\citep{sturner99}.  None of these explanations is straightforward or
without problems.  A firm measurement of the \Al gamma-ray line width
is desirable, before questioning our understanding of \Al fate from
its production sites until decay in interstellar space.

New measurements have been obtained recently with Ge spectrometers
aboard the high energy solar spectroscopy imager RHESSI
\citep{smith03} and the INTEGRAL observatory. The RHESSI result was
derived through Earth occultation analysis of data while pointing at
the sun. They obtain an intermediate
intrinsic line width of about $(2.0 \pm 0.8)$~keV FWHM. The
preliminary INTEGRAL measurement also suggests a narrow line, but
systematic uncertainties are still large \citep{diehl03}. Within
the given uncertainties, the current set of measurements is mildly
inconsistent. But more measurements have been recorded with INTEGRAL's SPI
Ge spectrometer already, so that longitude-resolved line width results
are at the horizon.

We explore the potential of \Al decay line shape measurements for
diagnostics on the sources of \Al and their interstellar environment.
We follow up on earlier analysis by \citet{gehrels96}, who
showed that the structure of the Galaxy should be reflected in the
position of the line due to Doppler shift from Galactic rotation. In
this paper we employ an updated spatial distribution model for the
plausible sources of \Al in the Galaxy.  We also account for the fate
of ejected \Al in interstellar space through different model variants
which reflect ejections by winds and/or supernovae into cavities of
plausible sizes for the massive-star environment of the \Al sources.
Our aim is to illustrate the diagnostic power of measuring the \Al
gamma-ray line's position and width; this should be feasible with
current instrumentation through imaging spectroscopy, even if a
detailed decomposition of the line shape may still be beyond
reach.

\section{\Al Sources in the Galaxy}

Our model for the distribution of \Al in the Galaxy is based on two
separate aspects: the distribution of nucleosynthesis events in the
Galaxy, and the distribution of \Al in interstellar space following an
individual nucleosynthesis event.  For each of these, spatial as well
as velocity-space densities have to be considered.

\begin{figure}[t]
  \centering
  \includegraphics{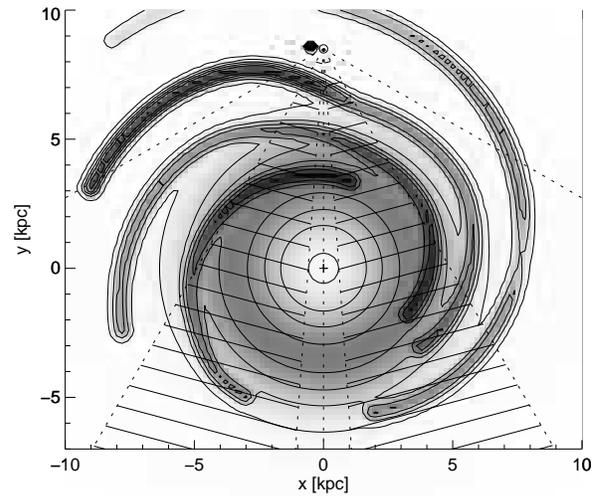}
  \caption{Model of the free electron density in the galactic plane
    \citep{taylor93}, used as \Al source density distribution. Dotted
    lines represent $-60\degr$, $-30\degr$, $-4\degr$, 0\degr,
    4\degr, 30\degr and 60\degr\ galactic longitude. Hatched
    areas illustrate the eastern/western parts of the inner galactic
    region.}
  \label{fig-tc-model}
\end{figure}

The angular distribution of \Al 1809~keV emission on the sky
correlates well with tracers of ionisation, such as H$_\alpha$ or
free-free emission. Therefore we adopt a three-dimensional model for
the space density of free electrons as our parent distribution for \Al
sources in the Galaxy. Such a model has been derived by
\citet{taylor93} from pulsar dispersion measure observations (a cut
through this model along the Galactic plane can be seen in
Fig.~\ref{fig-tc-model}); it was updated in \citet{cordes02}.
Alternative spatial models, such as the smooth, axisymmetric one by
\citet{gomez01} will probably yield similar line shape results.

The Doppler shifts due to Galactic rotation can then be determined
from the Galactic rotation curve. For our model we 
used the results obtained by \citet{olling00} from fitting radial
velocity measurements with a
five-component mass model of the Galaxy, consisting of a stellar
bulge, a stellar disc, two gas discs (\ion{H}{i}, H$_2$) and a
dark-matter halo. Because their determination of the rotation curve
depends on the values of the distance to the galactic centre $R_0$ and
the local circular speed $\Theta_0$, which are not known to a high
degree of precision, they allowed these parameters to vary over a
broad range of values.  In view of these uncertainties and because we
are interested mainly in the inner Galaxy, we approximated the
rotation curve given by Olling \& Merrifield for the IAU standard
values $R_0=8.5\ \mathrm{kpc}$ and $\Theta_0=220\ \mathrm{km~s}^{-1}$
with the radial dependence:
\[ |\vec v|(R) = 220\ \mathrm{km~s}^{-1} \cdot 
\left[1 - \exp(-R / 1\ \mathrm{kpc})\right] \]
the velocity vector being parallel to the plane and perpendicular
to the vector pointing from the galactic centre to the source location.

Superimposed on Galactic rotation is the motion of freshly synthesised
radioactive material due to the parental supernova explosion or the
ejecting Wolf-Rayet star wind and its slowed-down motion in the ISM
before decay, i.e.\ within $\approx 10^6$~yr.
Concentrating first on supernovae, we
adopt a particular expansion behaviour: Recent hydrodynamic
simulations of type~II supernovae by \citet{kifonidis03} find that the
expansion of the bulk of nucleosynthetic SN products such as \Al may
be at velocities less than 1200~km~s$^{-1}$. To reflect their results,
we allow \Al to expand freely with a velocity of 1500~km~s$^{-1}$
until it reaches the radius of the SN reverse shock formed by
circumstellar interaction. After this point, we expand \Al at the
velocity of the blast wave shock; this gives us a conservative
estimate because \Al is likely to move slower than the forward shock.
For our model of SNR dynamics from
circumstellar interaction, we adopt the values for Kepler's supernova
remnant shock positions and velocities given by \citet{mckee95}. 

At present, our model does not include other \Al sources; Type Ib/Ic
supernovae and Wolf-Rayet stars \citep{prantzos96}, which eject matter
at similar or even higher speeds than Type II SNe
\citep{mellema02,prinja90,garcia96}. Also, the interaction of ejected
matter with the surrounding medium depends on the star formation
history of the source region, where bubbles forming around groups of
young massive stars play a potentially large role.  Cavities extending
over several hundred pc have been observed in galaxies \citep{oey96},
and the Eridanus cavity \citep{burrows93} presents us with a nearby
example of such a cavity, extending from the Orion star forming region
to very near the Sun. Matter ejected into such a low-density bubble
could expand almost freely until reaching the boundary whereas
typically assumed ISM densities $(n_\mathrm{H}\simeq
1~\mathrm{cm}^{-3})$ would slow it down quite rapidly. To keep the
model simple at first, we adopt above SNR model as typical for \Al
sources; refinements will be discussed in our subsequent studies.

With these assumptions, the ejected \Al moves freely for $\approx
2$~kyr, then decelerates with the SNR's shell. The shell reaches a
radius where it dissolves in the ISM at an age comparable to the
lifetime of \Al, when a significant fraction has therefore already
decayed.  We note that the expansion velocity drops below the
characteristic rotational velocity of 220~km~s$^{-1}$ at $\approx
40$~kyr, when 96\% of the \Al is still left. Therefore the
contribution from expansion to the overall line width will be rather
small in our model.

\section{Line Shape Diagnostics}

We obtain simulated sky maps and spectra from a Monte Carlo scheme:
\Al source locations are chosen randomly from a spatial distribution
proportional to the free electron density, distributing candidate
source positions within a volume centered on the Galaxy and extending
25~kpc in the plane and 7.5~kpc perpendicular to the plane. The
free-electron density within this volume is taken from
\citet{taylor93}. For each nucleosynthesis event, a random 
age is chosen within the interval $[0, 10^7\ \mathrm{yr}]$. From this we
evaluate the intrinsic velocity distribution and size of the 
\Al source, following the above expansion model.  The age of the
nucleosynthesis event thus determines extent and intrinsic velocity of
its ejecta, as well as their 1.809~MeV luminosity. We represent each
event by $2^{10}$ mass elements to reflect its spatial extent. 

With the observer at $R_0=8.5\ 
\mathrm{kpc}$ and $\Theta_0=220\ \mathrm{km\ s^{-1}}$, we obtain
viewing direction and radial velocity of \Al sources. Direction
and radial velocity give us the coordinates of the \Al source mass
element in a data space of \Al decay luminosity as a function of
longitude, latitude and photon energy. 

\begin{figure}[t]
  \centering
  \includegraphics{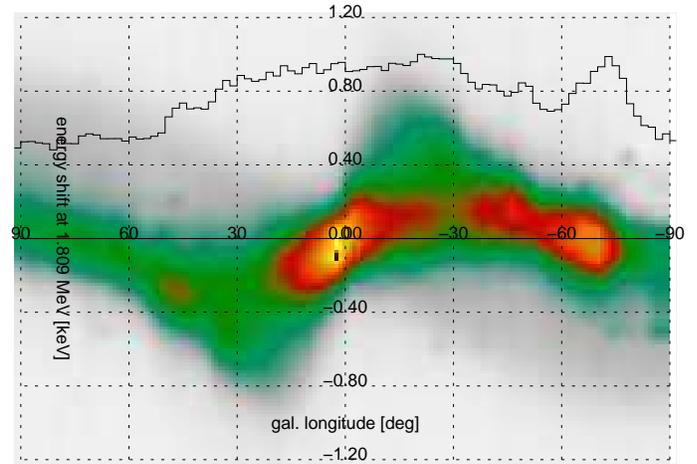}
  \caption{\Al line intensity as a function of Galactic longitude and
    gamma-ray photon energy, with the source longitude profile superimposed}
  \label{fig-long-spec}
\end{figure}

Projecting this data volume onto the longitude-energy plane, we obtain
a map of \Al line intensities as a function of Galactic longitude and
observed photon energy (See Fig.~\ref{fig-long-spec}). The intrinsic
velocity spreads and spatial source distributions lead to a line
blurring of $\approx 0.3-0.5$~keV along the plane of the Galaxy.
Galactic
rotation leads to the symmetric line shifts east and west of the
Galactic centre. Deviating from the trend of decreasing surface
brightness with increasing separation from the Galactic centre, we
notice a prominent flux enhancement around
$-70\degr$, which corresponds to the direction tangential to the
Carina-Sagittarius spiral arm, which is also clearly visible in the
longitude profile of our source distribution model, superimposed as a
histogram in the upper half of the figure; see also
Fig.~\ref{fig-tc-model}. 

\begin{figure}[t]
  \centering
  \includegraphics{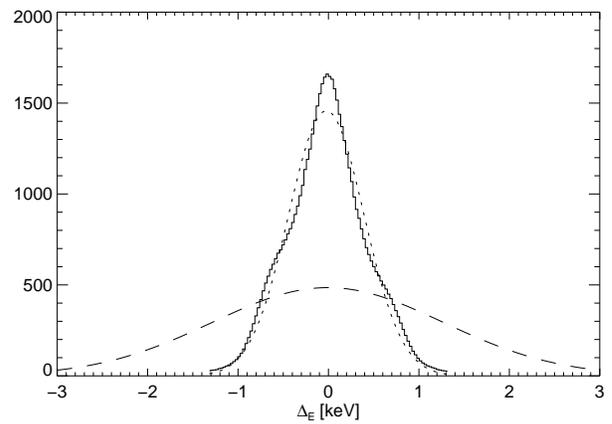}
  \caption{Spectrum of the inner galaxy ($l \in [-30\degr,
    30\degr]$. The full line is the source spectrum, the dotted line
    the best-fitting Gaussian, and the dashed line results from a
    convolution with a 2.8~keV FWHM Gaussian adopted for instrumental
        resolution of the measuring detector.
    $b \in [-5\degr, 5\degr]$)}
  \label{fig-in-gal-spec}
\end{figure}

In order to compare with a measured line profile, we integrate our
simulated skymap of line energies and intensities over a region of
interest of and obtain a resulting spectrum corresponding to an
observation of this region without spatial, but with perfect spectral
resolution.  The spectrum shown in Fig.~\ref{fig-in-gal-spec}
represents a region-of-interest chosen to reflect the RHESSI analysis
of the inner Galaxy \citep{smith03}.
Fitting a Gaussian to this spectrum yields an equivalent line width of
1.0~keV (FWHM). The deviation from a Gaussian shape, which appears
clear in our model spectrum (Fig.~\ref{fig-in-gal-spec}, continuous
line versus dotted line) is probably too small to be detectable with a
realistic Ge detector of instrumental resolution 2.8~keV FWHM
\citep[approximate for SPI on INTEGRAL, determined by in-orbit
measurements of detector background lines;][]{attie03}, convolution with
the instrument response suppresses the difference between the measured
flux and a Gaussian by a factor of $\approx 2000$ (see dashed line in
Fig.~\ref{fig-in-gal-spec}).

\begin{figure}[t]
  \centering
  \includegraphics{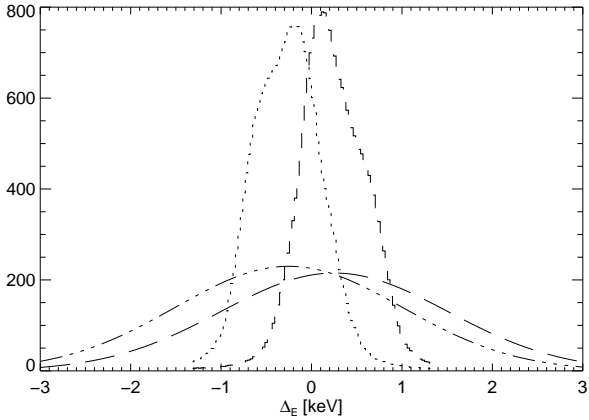}
  \caption{Illustrations of lines from different Galactic regions:
    The histograms (dotted/dashed lines) show the spectra of the
    eastern/western part of the inner galactic region (i.e.\ $\pm
    5\degr$ in latitude, $\pm [4\degr-30\degr]$ in longitude)
    Energy-resolution limited spectra (obtained by convolving with
    2.8~keV FWHM Gaussians) are shown as dash-dotted/long-dashed
    lines.}
  \label{fig-east-west}
\end{figure}
Nevertheless, spatially resolved observations can -- at least
partially -- separate the rotational effect from the total line
broadening. Fig.~\ref{fig-east-west} compares the model spectra of the
east and west part of the inner galactic region. (defined by $b \in
[-5\degr , 5\degr]$ and $l \in [4\degr, 30\degr]$, $l \in
[-30\degr, -4\degr]$, respectively) If we assume an instrument
with a spectral resolution of 2.8~keV FWHM, the observed spectra are
again almost indistinguishable from Gaussians, but we expect their
centroids to be shifted by a significant energy offset. For our choice
of parameters, she shifts amount to $-0.25$~keV for the eastern and
$+0.25$~keV for the western part. A real measurement would yield a
line width that is lower than the width obtained for the entire inner
Galaxy. When we fit Gaussians to these resolution-limited spectra, we
obtain $\approx 0.9$~keV, compared to $\approx 1.0$~keV for the whole
inner region ($b \in [-5\degr , 5\degr]$, $l \in [-30\degr,
30\degr]$, Fig.~\ref{fig-in-gal-spec}). This is due to the fact that
partial-region spectra only include a lower dynamic range of velocities.

We used data analysis parameters from the first inner-Galaxy
observations of INTEGRAL's core programme to estimate the possible
precision of gamma-ray line centroid measurements. By using Gaussian
fits to simulated spectra with varying levels of statistical noise,
we tested the relation of the achievable line position determination
accuracy and the signal-to-noise ratio. We confirm the centroid
uncertainty to be proportional to the noise-to-signal ratio, therefore
scaling with the inverse square root of exposure time. At present,
noise is dominated by systematic uncertainties, resulting in a position
uncertainty of $\pm 0.19$~keV.  If we extrapolate to the full exposure of 3~Ms
scheduled for the first year of INTEGRAL operations, we estimate a
statistical uncertainty for the \Al line centroid of $\pm 0.06$~keV.
We expect that systematic uncertainties in INTEGRAL results can be
reduced in the near future through studies of SPI's spectral
resolution, energy calibration stability, and background in-flight
behaviour. When systematic uncertainties are reduced below the
level of statistical uncertainty, we expect a $3\sigma$ detection of
the Doppler shift relative to an equally long observation at the
Galactic longitude of maximum Doppler shift at $l=330\degr$.

\section{Conclusions}

We model \Al sources in the Galaxy, adopting massive stars as the
dominating sources. Using Galactic rotation and plausible assumptions
for the fate of \Al from ejection by the supernova or Wolf Rayet star
into until decay in interstellar space, we derive an expected profile
for the 1809~keV gamma-ray line from the decay of Galactic
\Aln. For this, we integrate along the lines of sight over the
Doppler-shifted source regions of \Al emission with their intrinsic
state of dynamical evolution.

Measurements of the 1809~keV emission from galactic
\Al play in principle a role similar to the \ion{H}{i} 21~cm
radiation, with the benefit that the Galaxy is always optically thin
to MeV gamma radiation. Current instruments are unable to improve the
knowledge of galactic rotation over \ion{H}{i} results, the basis for
the inner galaxy rotation curve by \citet{olling00}, but the knowledge
of rotation can be used to test the position of \Al sources. 

Our result demonstrates that the gamma-ray line profile reflects the
kinematics of decaying \Al in the ISM. However, current Ge detectors
will probably be
unable to detect line shape departures from a simple Gaussian shape
for spectra such as predicted by our model. Nevertheless, centroid and
width of the line will provide a diagnostic for the \Al sources in the Galaxy.

Detection of the expected amount of line shift would put a limit on
the contribution of local emission to the \Al flux in the direction of
the inner galaxy. By their massive star origin, \Al and
\ion{H}{ii}-regions are connected; the latter having also been used
to measure galactic rotation \citep{brand93}. \ion{H}{ii} regions are
created by the ionising radiation from O stars, therefore probe the
first 2~My of star formation. The peak of \Al emission from a coeval
group of massive stars only begins at 2~My, continuing for another
10~My \citep{plueschke01}. This demonstrates that \Al and the other
analyses of galactic rotation may complement each other.

Our model
parameters, chosen to represent supernova-produced \Al with ejection
into an typical ISM environment, predict a line width of
$\approx$~1~keV for the inner Galaxy. This value is
on the low side of the recent RHESSI measurement \citep{smith03},
and significantly below the ``broad'' line reported from the GRIS
balloon measurement \citep{naya96}.  We suggest, therefore, that \Al
decay in the interstellar medium is likely to teach us more about the
ejection and expansion characteristics of nucleosynthesis ejecta, and
thus about the morphology of the interstellar medium in the vicinity
of massive stars, our assumed sources of \Al in the Galaxy. For
example, if \Al sources typically are
surrounded by cavities from earlier massive-star action, this could lead to
some additional broadening of the observed \Al gamma-ray line.
Further imaging spectroscopy analysis of RHESSI data and of
measurements with the SPI Ge spectrometer on INTEGRAL \citep[launched in
October 2002;][]{winkler03} promise to provide the data for such
studies.

\end{document}